\documentstyle[12pt,aasms4]{article}

\lefthead{H. Casini and R. Montemayor}
\righthead{Crust-core interactions and the magnetic dipole orientation
in neutron stars}

\begin{document}
\title{Crust-core interactions and the magnetic dipole orientation in neutron stars.}
\author{H.Casini and R.Montemayor}
\affil{Instituto Balseiro, Universidad Nacional de Cuyo\\
and \\
Centro At\'{o}mico Bariloche, CNEA\\
8400 S.C. de Bariloche, R\'{\i}o Negro, Argentina}

\begin{abstract}
We develop an effective model for a neutron star with a magnetosphere. It
takes into account the electromagnetic torques acting on the magnetic dipole,
the friction forces between the crust and the core, and the gravitational
corrections. Anomalous electromagnetic torques, usually neglected in a rigid
star model, play here a crucial role for the alignement of the magnetic
dipole. The crust-core coupling time implied by the model is
consistent with the observational data and other theoretical estimations.
This model describes the main features of the behavior of the
magnetic dipole during the life of the star, and in particular gives a
natural explanation for the $n<3$ value of the breaking index in a young
neutron star.
\end{abstract}


\section{Introduction}

Pulsars are identified with rapidly rotating, highly magnetized neutron
stars. The basic information we receive from this kind of stars is a sequence
of electromagnetic pulses with a very stable frequency, which is interpreted
as directly related to the star rotation. The angular velocity $\Omega$
generally decreases gradually as a result of the torque exerted on the star
by the radiation reaction. In the vacuum dipole model the star magnetic field
is assumed to be a magnetic dipole $M$ that forms an angle $\theta$ with the
rotation axis, so that the star loses energy by electromagnetic radiation due
to its rotation (\cite{P}; \cite{GO}). This turns out to be the main source
for the star energy loss. In this simple model the evolution of the angular
velocity for a neutron star with a momentum of inertia $I$ is
\begin{equation}
\,\dot{\Omega}=-\frac 23\frac{M^2\Omega ^3}{Ic^3}\sin {}^2\theta \,\,\,,
\label{1}
\end{equation}
or, more generally, $\,\dot{\Omega}=-\mu \Omega ^3$. Several
characteristics of the dynamics of the $\mu $ parameter have been observed.
After glitches a sudden increment of $\mu $ has been noticed, which does not
completely relax back (\cite{LEB}; \cite{LEa}). There is also evidence that
for old pulsars, with an age of $\frac \Omega {2\dot{\Omega}}\sim 10^7yr$,
$\mu$ is smaller than for younger pulsars (\cite{RZC}). The first behavior
can be interpreted as an increase of the external torque after de glitches,
whereas the second one suggests a slow decrease of the torque with the age of
the star.

Besides this, the breaking index $n=\frac{\Omega \ddot{\Omega}}
{\dot{ \Omega}^2}$ has been measured in four young pulsars, and it takes
values between $1.4$ and $2.8$. If $\mu$ were constant the breaking index
would be exactly equal to $3$, but the measured values are smaller than this
canonical value implying an increasing $\mu$.
A number of factors might affect the breaking index. One of them is the
presence of mechanisms of loss of energy different from the dipolar
electromagnetic radiation that could change the $\Omega$ exponent in eq.
(\ref{1}). For example, pure multipolar electromagnetic radiation gives
$n\geq 5$ , gravitational quadrupole radiation $n=5$ (\cite{MT}), and early
neutrino emission $n<0$ (\cite{AO}). These effects are expected to be
relatively weak for the pulsars with a measured breaking index, and in
particular the first two would increase $n$ and thus are unable to explain
the observed values smaller than $3$. Another possible factor that could be
considered is the effect of the pulsar magnetosphere (\cite{GJ}), but
although it would change appreciably the angular dependence and the
numerical coefficients in eq. (\ref{1}), it leaves the $\Omega$ exponent
equal to $3$ (\cite{GN}), and thus does not affect the value of the breaking
index provided that $\mu$ remains constant (see references (\cite{M}) for an
alternative).  According to this, the most natural explanation seems to
require variations of the parameters $I,M$ and $\theta$, which change the
breaking index in the following way for a star without magnetosphere:
\begin{equation}
n=3+\frac \Omega {\dot{\Omega}}\frac{\dot{\mu}}\mu =3+\frac \Omega {\dot{
\Omega}}\left( 2\frac{\dot{\theta}}{\tan \theta }+2\frac{\dot{M}}M-\frac{
\dot{I}}I\right) \;.  \label{2}
\end{equation}
Given that $(n-3)$ is of the order of the unity, the parameters responsible
for the variation should have characteristic evolution times of the order of
the pulsar age $\tau _\Omega =\frac \Omega {2\dot{\Omega}}$, which is of
around $1000yr$ in the case of the stars of known breaking index.

With respect to the momentum of inertia, sensible variations during such
times due to changes in the structure, sphericity, or in the coupling of the
internal superfluids to the crust, seem to be unable to explain the observed
deviations (\cite{LEa}).

Significant changes in the magnetic moment also seem to be inhibited by the
high crust conductivity, which can keep the magnetic fields unchanged and
thus fixed to the crust along times of order $10^7yr$ or more (\cite{L};
\cite{Ch}; \cite{PK}; \cite{GR}) and hence the breaking index $n$ is
unaffected in young pulsars. Furthermore, the decay of the magnetic field
reduces the radiation rate and hence would lead to $n>3$. However, there is
no general consensus about the behavior of the intensity of the effective
magnetic moment and other hypotheses have been proposed. Muslinov and Page
(\cite{MP}) consider magnetic fields trapped under the surface during the 
birth of the star, which could rapidly be freed and would thus increase the 
total magnetic moment (see also (\cite{BAH}; \cite{ChS})). Ruderman 
(\cite{R}) considers the possibility of crust cracking, reordering due to 
vortex pinning and stresses. If flux tubes in the core proton superconductor 
are pinned to neutron superfluid vortices, the latter could drag the 
magnetic structure to the equator as the star spins down (\cite{RZC}; 
\cite{SBMT}).  In this work we adopt the first hypothesis and simply assume 
a stable external magnetic field rigidly fixed to the crust, except perhaps 
in very short periods at the glitches.

Under the above assumptions, the only responsible for the anomalous breaking
index could be the variation of the angle between the magnetic dipole moment
${\bf M}$ and the rotation axis ${\bf \hat{\Omega}}$. In this case eq.
(\ref{2}) becomes:
\begin{equation}
n-3=2\frac \Omega {\dot{\Omega}}\frac{\dot{\theta}}{\tan \theta }\;.
\end{equation}
The angular momentum of the star decreases during its life by radiation, and
thus $\dot{\Omega}$ is negative. Therefore a value $n<3$ requires an increase
in time of $\theta$, e.g. $\dot{\theta}>0$. In brief, for a rigid star in
vacuum the observed dynamics is related to the behavior of the $\theta$
angle. The observational data suggest that its evolution is controlled by
three characteristic times. There is a fast dynamics, associated to the
reorientation of the magnetic dipole during the glitches, with a
characteristic time less than or of the order of 50 days, an intermediate
dynamics that dominates during the early stages of the stars, where the
magnetic moment slides towards the equator and gives the breaking index
values smaller than $3$, and finally a slow dynamics, where the magnetic
moment slides towards the rotation axis of the star, which involves
characteristic times of the order of $\tau _\Omega \sim 10^7yr$ (\cite{LM}).
The intermediate dynamics competes with the slow one, and the observational
evidence shows that for young pulsars it dominates, leading to the anomalous
values of the breaking index, whereas for old pulsars ($\tau _\Omega \sim
10^7$ years) it is the slow one which dominates, giving place to the
alignment of the magnetic dipole with the rotation axis. However, if we only
took into account the radiation reaction of the magnetic dipole on a rigid
star, the evolution of $\theta$ would perform a slow alignement during the
life of the star (\cite{PP}) with a breaking index $n>3$, in contradiction
with the features mentioned before.

In this work we consider a more realistic description given by a simple
effective model, where the star is considered as constituted by two rigid
interacting components in presence of a magnetosphere. One of them is the
core which contains the bulk of the mass, with a momentum of inertia $I_o$
and an angular velocity ${\bf \Omega }_o$. Its angular momentum is mainly
given by the vortices of superfluid neutrons. The other component is the
crust, with a momentum of inertia $I_c\ll I_o$ and an angular velocity ${\bf
\Omega }_c$, and with a dipole magnetic moment ${\bf M}$. The evolution of
the system is governed by three kinds of torques:

a.- torques acting on the crust due to electromagnetic interactions,

b.- friction torques between the core and the crust due to the interaction
of the neutron vortices, the proton flux tubes and the electrons in the star,

c.-torques due to gravitational effects.

Although this model could seem a rather simplified approach, it is sufficient
for obtaining a good qualitative comprehension and can be considered as a
first step to a more sophisticated description.

In the following section we present a review of the different effective
torques we take into account in our model. In Section III we write the system
of equations that describes the dynamics of the model and we analyze the
resulting behavior. Finally, the last two sections are devoted to the
discussion of the results and their phenomenological implications.

\section{The effective torques}

In this section we review the torques that act on both components of the
star. In principle the coordinates of our model are represented by three
vectors, ${\bf M}$, ${\bf \Omega}_o$ and ${\bf \Omega}_c,$ which are the
magnetic moment, the core angular velocity and the crust angular velocity
respectively. It should be remarked that the two components considered in the
present work are identified with the crust and core, and they are not the
ones used to describe the long relaxation time after glitches, which is
currently identified as an effect of the crustal superfluid (\cite{AI};
\cite{ACP}; \cite{LEb}). Once we have chosen a direction of reference there
remains five degrees of freedom, because we have assumed the magnetic dipole
moment constant in modulus and fixed to the crust. So the equation of motion
for the magnetic field is simply
\begin{equation}
{\bf {\dot{M}=\Omega }}_c{\bf {\times M\;.}}
\end{equation}
In the following subsections we describe the torques that arise from the
different interactions.

\subsection{Electromagnetic torques on the magnetic dipole}

We have a magnetic dipole ${\bf M}$ which forms an angle $\theta$ with the
angular velocity ${\bf {\Omega}}_c$ of the crust. There are several dipole
and quadrupole torque terms that act on the magnetic dipole, and as we have
assumed that it is bound to the crust, these torques are directly applied to
the latter. They have been carefully discussed by Good and Ng (1985). The
effect of the average torques that involve quadrupole terms are strongly
suppressed, and hence it is enough to consider only the dipolar components.
There are two types of dipolar torques that act on the crust, the
non-anomalous torques of order $\frac{M^2\Omega _c^3}{\,c^3}$, which are of
the same order as the classical spin down torque (\cite{DG}; \cite{MG}), and
the anomalous torques that are $\frac c{\Omega _cR}$ times larger than the
former ones, where $R$ is the radius of the star. In the following we will
use an orthogonal system of reference with ${\bf \hat{z}}$ in the ${\bf
\Omega}_c$ direction, ${\bf \hat{y}}$ in the plane $({\bf \Omega}_c,{\bf
M})$, and ${\bf \hat{x}}$ orthogonal to such a plane, according to Fig. 1.
The ${\bf \hat{y}}$, and ${\bf \hat{z}}$ torques, which cause the loss of
energy and the alignment of the magnetic dipole in a rigid star, are non
anomalous. In this plane the torque is

\begin{equation}
{\bf T}_{yz}=I_c\,\omega _{yz}{\bf \,\hat{{M}}\times (\hat{M}\,{\ \times }{
\Omega }_c{)-}}I_c\tilde{\omega}_{yz}{\bf {\Omega }_c\;,}
\end{equation}
where ${\bf \hat{M}}=M^{-1}{\bf M}$, $\omega _{yz}=\frac 23\frac{M^2\Omega
_c^2}{I_c\,c^3}\nu _{yz}$ and $\tilde{\omega}_{yz}=\frac 23\frac{M^2\Omega
_c^2}{I_c\,c^3}\tilde{\nu}_{yz}$. The $\nu _{yz}$ and $\tilde{\nu}_{yz}$
coefficients contain information on the magnetosphere contribution to the
torque. The first term changes the magnitude and direction of ${\bf{\Omega}
_c}$, whereas the last one alters only its magnitude. For a neutron star
without magnetosphere we have $\nu _{yz}=1$ and $\tilde{\nu}_{yz}=0$, but the
presence of the magnetosphere may greatly change the value of these
coefficients. Its effects are in general of comparable size to the ones in
the vacuum case, and could depend on the angle between ${\bf \hat{M}}$ and $
{\bf {\Omega}}_c$. Their values are related not only to the currents flowing
in the near magnetosphere, but also to the ones in distant regions, and so
cannot be calculated at present. For this reason we will maintain these
coefficients of the order of the unity as unknown adimensional functions of
the angle between the dipole magnetic moment and the angular velocity, to be
phenomenologically estimated.

The ${\bf \hat{x}}$ component has anomalous and normal terms, but the first
ones dominate because the normal terms are much smaller than the anomalous
ones. The anomalous contribution is given by
\begin{equation}
{\bf T}_x=\,I_c\,\,\omega _x \,\cos\theta {\bf \,\;{\Omega }}_c{\bf {
\times \hat{M}\;,}}
\end{equation}
where $\omega _x=\frac 45\frac{M^2\Omega _c^2}{I_c\,c^3}\left( \frac c{
\Omega _cR}\right) \nu _x$. Here the $\nu _x$ coefficient contains
information on the magnetosphere, and reduces to $\nu _x=1$ when it is
absent. In the case of the anomalous torque the magnetospheric effects
depend mainly on the near region and can be evaluated under reasonable
assumptions. For example, if we assume that there is a dominant contribution
from the Goldreich-Julian charge density on the closed lines of forces, it
leads to $\nu _x=-1/4$.

As long as $\frac c{\Omega _cR}$ $\sim 10^2-10^4$ for most pulsars, $\omega
_x$ is much greater than $\omega _{yz}$ and $\tilde{\omega}_{yz}$. Despite
this, in the rigid dipole model the anomalous term does not contribute to
the alignment of the magnetic dipole with respect to ${\bf \Omega}_c$ or to
the spin down of the star. However, in the two-component model considered
here the situation changes and the anomalous term acquires a significant
role.

\subsection{Dissipative effects in the superfluid hydrodynamics}

In the outer-core region of a neutron star there is a mixture of superfluid
neutrons, superconducting protons and normal electrons. The effects of
viscosity and friction due to the scattering of the electrons by the neutron
and proton vortices give place to effective torques between the core and the
crust. The resulting crust-core friction can be characterized by a time
parameter $\tau _f$. There are several phenomenological estimations on the
basis of the glitch dynamics. For example, upper bounds are given for the
Vela pulsar of $\tau _f<10s$ for a crust initiated glitch and of $\tau
_f<440s$ for a core initiated glitch (\cite{AEO}). These values are deduced
from the December 24, 1980 Christmas glitch data (\cite{MHMK}).

This point has also awakened a great deal of theoretical interest. The
interior plasma couples the neutron superfluid due to mixing superfluid
effects (\cite{ALS}; \cite{AS}), and it is locked to the crust due to Alfven
waves or cyclotron vortex waves. An extensive analysis of these phenomena has
been performed by Mendell (\cite{M1}, 1997) on the basis of the Newtonian
superfluid hydrodynamics, generalized to include dissipation. The
characteristic times related to all these couplings are of the order of the
second. The mutual friction torque simplifies to
\begin{equation}
{\bf T}_f=f\left( {\bf \Omega }_c{\bf {-\Omega }}_o\right) \;,
\end{equation}
when the angular velocities differences are small. When the system is taken
out of equilibrium such as in a glitch, the time response frequency is given
by $\frac 1{\tau _f}=\omega _f=f\frac{I_t}{I_cI_o}$, where $I_c$ and $I_o$
are the moments of inertia of the components, and $I_t=I_c+I_n$ is the total
star moment of inertia. Therefore this torque is relevant for the rapid
nucleus spin up during glitches. The effective friction could depend on the
angular velocity. Non dissipative effects give torques of the same kind as
the gravitational dragging, which is to be discussed in the next subsection.
If they are also characterized by a time of the order of the second they
should be smaller than the gravitational torque for a rapidly rotating star.

\subsection{Gravitational effects}

The neutron star is essentially constituted by superfluid matter, and hence
the observed rotation can be achieved only by the presence of vortices. The
gravitational fields are strong enough to be relevant. They induce a change
in the shape of the vortex lines, and also affect the density of vortices
(\cite{CM}). The main contribution of these effects is a correction on the
coefficients of the dissipative torques of the order of 15\% with respect to
the flat space-time values, but they do not introduce new terms.

An additional term arises from the gravitational dragging, which gives place
to a new torque if the components of the star have different angular
velocities. If we neglect the gravitational radiation this interaction is
conservative and the corresponding torque is
\begin{equation}
{\bf T}_g=z{\bf \,\Omega }_c\times {\bf {\Omega }}_o\;,  \label{8}
\end{equation}
where $z\simeq I_cI_o\frac{2G}{c^2R^3}\simeq I_c\frac{R_s}R\left( \frac{R_g}R
\right) ^2\simeq 0.1$ $I_c$ for small velocities, $\Omega R\ll c$, where $R$
is the star radius and $R_g$ and $R_s$ are the gyration and Schwarzschild
radius of the star. This torque has a significant modulus, and at a first
glance it could have sensible effects on the dynamics of the magnetic dipole.

The torques ${\bf T}_{f}$ and ${\bf T}_{g}$ could be seen as the
first terms of the mutual torque in an expansion around the point of equal
angular velocities, and as such the form of the interaction is largely
independent of the model. We are only considering terms up to the first
order in the angle between ${\bf \Omega}_{c}$ and ${\bf {\Omega}}_{o}$. As
we will discuss later this is enough for our analysis. In the following
section we will use all these torques to construct a set of equations of
motion that defines the dynamics of the star components.

\section{The equations of motion}

The equations of motion for the system described in the preceding section
are:
\begin{eqnarray}
{\bf {\dot{M}}} &=&{\bf {\Omega }}_c\times {\bf {M\;,}}  \label{9} \\
{\bf {\dot{\Omega}}}_c &=&\omega _x\cos \theta {\bf {\,\Omega }}_c{\bf {
\times }\hat{{M}}+}\omega _{yz}{\bf \,\hat{{M}}}\times {\bf (\hat{{M}}\,{\ }}
\times {\bf {\Omega }}_c{\bf {)-}}\tilde{\omega}_{yz}{\bf {\Omega }}_c{\bf -}
\frac f{I_c}({\bf {\Omega }_c{-}\Omega }_o){\bf -}\frac{z\,}{I_c}{\bf
\,\Omega }_c\times {\bf {\Omega }}_o\;, \\
{\bf {\dot{\Omega}}_o^{\prime }} &=&\frac f{I_o}\left( {\bf {\Omega }}_c{\bf
{-\Omega }}_o\right) {\bf +}\frac z{I_o}{\bf \,\,\Omega }_c\times {\bf {
\Omega }}_o\;.
\end{eqnarray}
If $f\rightarrow 0$ the crust decouples from the core and if $f\rightarrow
\infty $ the two components act as a rigid body. In both cases we recover
the dynamics of the model of a rigid star with a magnetic dipole, with
momentum of inertia $I_c$ and $I_t=I_c+I_o$ respectively, which has been
already discussed in the Introduction.

The eq. (\ref{9}) implies that the magnitude of ${\bf {M}}$ is constant.
Therefore we have only five variables in the system, three angles, $\alpha
,\,\beta ,\,$and $\theta $, defined in Fig. 1, and the two moduli of the
angular velocities, $\Omega _c$ and $\Omega _o$ . For these variables the
equations of motion are:
\begin{eqnarray}
\dot{\Omega}_c&=&-\frac f{I_c}(\Omega _c-\Omega _o\cos \,\alpha )-\omega
_{yz}\Omega _c\,\sin ^2\theta {-}\tilde{\omega}_{yz}{\Omega }_c\;,
\label{12} \\
\dot{\Omega}_o&=&\frac f{I_o}(\Omega _c\cos \,\alpha -\Omega _o)\;,
\label{13} \\
\frac{d(\cos \theta )}{dt}&=&\frac f{I_c}\frac{\Omega _o}{\Omega _c}\,(\cos
\beta -\cos \alpha \cos \theta )+\omega _{yz}\,\sin ^2\theta \,\cos \theta +
\frac{z\,}{I_c}\Omega _o\,\left( {\bf {\hat{\Omega}}}_o{\bf .}\left( {\bf {
\hat{\Omega}}}_c{\bf {\times }{\hat{M}}}\right) \right) \;, \\
\frac{d(\cos \,\beta )}{dt}&=&\frac f{I_o}\,\frac{\Omega _c}{\Omega _o}(\cos
\theta -\cos \beta \cos \alpha )+\Omega _c\left( 1-\frac z{I_o}\right)
\,\left( {\bf {\hat{\Omega}}}_o{\bf .}\left( {\bf {\hat{\Omega}}}_c{\bf {
\times }{\hat{M}}}\right) \right) \;, \\
\dot{\alpha}&=&-\frac{\,\omega _x\cos \theta }{\sin \alpha }\left( {\bf {
\hat{\Omega}}}_o{\bf .}\left( {\bf {\hat{\Omega}}}_c{\bf {\times }{\hat{M}}}
\right) \right) {\bf -}\sin \alpha \left(\frac f{I_c}\frac{\Omega _o}{
\Omega_c}{\bf +}\frac f{I_o}\,\frac{\Omega_c}{\Omega _o}\right) \nonumber\\
&&{\bf -}\frac{ \omega _{yz}\cos \theta }{\sin \alpha }(\cos \beta -\cos
\alpha \cos \theta )\;.
\end{eqnarray}

In principle this system of equations would be very difficult to solve, but
in fact it contains several dynamics with very different time scales, given
by $\omega _{yz}\sim\tilde{\omega}_{yz}\ll \omega_x\ll \min (\omega
_f\simeq\frac f{I_c},\Omega_c)$, which greatly simplify its treatment as we
will see now. In the first place, from eqs. (\ref{12}) and (\ref{13}), if
$\Omega_c$ and $\Omega_o$ are very different at a given instant, they will
attain equilibrium in a relatively short time of the order of $\tau _f$. The
last equation tells us that the transient of $\alpha$ is also characterized
by $\tau _f$, and thus, after a time of this order, this variable will
acquire a value of the order $\frac{\omega_x}{\omega_f}\ll 1$. Hence, after
this transient we will have $\Omega_c\sim \Omega _o\,$, $ \alpha \ll 1$ and
therefore $\theta \sim \beta$, all of them satisfying a slow dynamics with
characteristic frequencies of the order of $\omega_x$ or $\omega _{yz}$.

As was already commented, the first two equations imply that the moduli of
the angular velocities of the crust and the core will rapidly reach an
equilibrium regime where $\dot{\Omega}_c\simeq \dot{\Omega}_o$ and $\Omega
_c-\Omega _o\simeq -\left( \tilde{\omega}_{yz}+\omega _{yz}\sin ^2\theta
\right) \,\frac \Omega {\omega _f}$ , i.e., the difference between the
angular velocities is quickly suppressed. The common decrease of both
components is given by
\begin{equation}
\dot{\Omega}_o\simeq \dot{\Omega}_c\simeq -\frac{I_c}{I_t}\left( \tilde{
\omega}_{yz}+\omega _{yz}\,\sin ^2\theta \right) \Omega _c\;,
\end{equation}
which coincides with the one obtained in the dipole model for a rigid
star when there is a magnetosphere. This means that the torque component
along $x$ will not affect the decrease of the angular velocity. This is
analogous to the case of a rigid star, where this torque does not cause any
spin-down. It means that the dissipative effects of the core-crust
interactions do not change the star energy significantly, and therefore the
main mechanism of energy loss is still due to the electromagnetic radiation.
In fact, the ratio between the power lost by friction and radiation is
$\frac{W_f}{W_r}\simeq\frac{\omega_{yz}}{\,\omega_f}\ll 1$. However, as we
will see, $T_x$ produces a significant effect, because it makes an important
contribution to the orientation of the magnetic axis.

Returning now to the angular variables, we can use a first order
approximation in $\alpha$ and consider the quasi-stationary regime. To
simplify the expressions, instead of dealing with the very similar variables
$\theta$ and $\beta$, we will replace the last angle by a new angle $ \gamma$
defined as in Fig. 1. It satisfies $\cos\beta =\cos\alpha\cos
\theta+\sin\alpha\sin\theta\cos\gamma$, and we have
${\bf\hat\Omega}_o{\bf\cdot}
{\bf {\hat{\Omega}}}_c{\bf{\times}{\hat{M}}}=\sin\gamma
\sin\alpha\sin\theta$. Thus, with this substitution and at first order in
$\alpha$, assuming $\frac{I_o}{I_t}\simeq 1$,
$\Omega_o\simeq\Omega_c=\Omega$, we have:
\begin{eqnarray}
\dot{\theta} &=&-\omega _f\alpha \cos \gamma -\omega _{yz}\sin \theta \cos
\theta -\frac z{I_c}\Omega \alpha \sin \gamma \;,  \label{18} \\
\dot{\alpha} &=&-\,\omega _x\cos \theta \sin \theta \,\sin \gamma -\omega
_f\alpha -\omega _{yz}\cos \gamma \sin \theta \cos \theta \;, \\
\dot{\gamma} &=&-\Omega \left( 1-\frac z{I_c}\right) -\frac{\,\omega _x}
\alpha \cos \gamma \sin \theta \cos \theta +\frac{\omega _{yz}}\alpha \sin
\theta \cos \theta \sin \gamma \;.
\end{eqnarray}
The solutions for $\alpha$ and $\gamma$ can be decomposed in a transient
dynamics, with a characteristic time $\tau _f$, plus a slow varying
time-function. This implies that $\alpha$ and $\gamma$ will reach a
quasi-stationary regime in a few seconds. From here on they will be driven by
the slow time-dependence of $\theta$. This assertion can be verified by
evaluating the Liapunov exponents at the equilibrium point whose real parts
are $-\omega_f$. The equation for $\dot{\theta}$ contains only small
frequencies. The explicit solutions at the fixed point for $\alpha$ and
$\gamma$ are:
\begin{eqnarray}
\sin \gamma &=&\pm \frac{\,\omega _f\omega _x-\left( 1-\frac z{I_c}\right)
\Omega \omega _{yz}}{\sqrt{\left( \omega _x^2+\omega _{yz}^2\right) \left(
\omega _f^2+\left( 1-\frac z{I_c}\right) ^2\Omega ^2\right) }}\;,  \label{21}
\\
\alpha &=&\mp \sqrt{\frac{\omega _x^2+\omega _{yz}^2}{\omega _f^2+\left( 1-
\frac z{I_c}\right) ^2\Omega ^2}}\sin \theta \cos \theta \;.  \label{22}
\end{eqnarray}

These results are consistent with our previous discussion. In particular $
\alpha$ becomes of order $\frac{\omega _x}{\max (\omega _f,\Omega )}$, and
thus the approximation $\alpha \ll 1$ is totally justified. We can also see
here that the gravitational dragging torque, despite its noticeable
magnitude according to eq. (\ref{8}), has the only effect of renormalizing
the angular velocity $\Omega $ by a correction of the order of $10\%$.
In what follows we use $\kappa = 1-\frac{z}{I_c}\sim 0.9$.

Substituting the expressions (\ref{21}) and (\ref{22}) into eq. (\ref{18}),
we finally obtain
\begin{equation}
\dot{\theta}=\frac 23\frac{M^2\Omega^2}{I_c\,c^3}\frac{\Omega^2}{\omega
_f^2+\kappa^2\Omega ^2}\left( \frac{6\nu _x\omega _fc}{5\Omega ^2R}-\kappa
\nu _{yz}\right)\sin \theta \,\cos \theta \;,  \label{23}
\end{equation}
and we have for the angular velocity of the star
\begin{equation}
\dot{\Omega}= -\frac 23\frac{M^2}{I_tc^3}\left( \nu _{yz}\sin ^2\theta +
\tilde{\nu}_{yz}\right) \Omega ^3\;.  \label{24}
\end{equation}
With these results, the expression for the breaking index becomes:
\begin{equation}
n=3\left[ 1+\frac 13\frac{I_t}{I_c}\frac{\Omega ^2}{\omega _f^2+\kappa^2
\Omega ^2}\left( \frac{6\nu _x\omega _fc}{5\Omega ^2R}-\kappa\nu _{yz}\right)
\sin \theta\,\cos \theta \;\frac d{d\theta }\left( \nu _{yz}\sin ^2\theta
+\tilde{\nu}_{yz}\right) ^{-1}\right] \;.  \label{25}
\end{equation}
The equations of motion (\ref{23}), (\ref{24}), and the expression (\ref{25})
for the breaking index show that the effects of the magnetosphere are
indeed relevant to explain the evolution of the magnetic dipole and the
angular velocity of a neutron star.

\section{Discussion}

The model depends on several parameters. The mechanical characterization of
the system is given by $I_t/I_c$ and $R$. The lower magnetosphere is
described by $\nu _x$, whereas the upper one is represented by $\nu _{yz}$
and $\tilde{\nu}_{yz}$. The effective friction between the core and the crust
is given by $\omega _f$. We have observational information on $\Omega$,
$\dot{\Omega}$ and $n$. Furthermore, we expect to have a star radius of the
order of $R\sim 10\;km$, a ratio of the total momentum of inertia and
the crust momentum of inertia $I_t/I_c\sim 10^2$ and $\kappa\sim 1$.
The angular velocities of neutron stars are in the range $1\;s^{-1}>\Omega
>10^3s^{-1}$, whereas a reasonable value for $\omega _f$ is of the order of
$1s^{-1},$ and thus we can assume that $\frac{\Omega^2}{\omega_f^2+\kappa^2
\Omega^2}\sim 1$.

The simplest situation we can consider corresponds to a star without a
magnetosphere. In this case $\nu _x=\nu _{yz}=1$ and $\tilde{\nu}_{yz}=0$,
and hence we have:
\begin{eqnarray}
\dot{\Omega} &=&-\frac 23\frac{M^2\Omega ^3}{I_tc^3}\sin ^2\theta \;, \\
\dot{\theta} &\simeq &\frac 23\frac{M^2\Omega ^2}{I_cc^3}\;\left( \frac{
6\omega _fc}{5\Omega ^2R}-1\right) \sin \theta \,\cos \theta \;, \\
n &\simeq &3\left[ 1-\frac 23\frac{I_t}{I_c}\left( \frac{6\omega _fc}{5\Omega
^2R}-1\right) \cot ^2\theta \right] \;.
\end{eqnarray}

The first equation states that $\dot{\Omega}$ is always negative, which is
consistent with the fact that the star is losing energy by electromagnetic
radiation, and thus the angular velocity is constantly decreasing. The
second equation implies that the magnetic dipole slides toward the direction
of the axis of rotation if $\frac{6\omega _fc}{5\Omega ^2R}<1$ or to the
equator if $\frac{6\omega _fc}{5\Omega ^2R}>1$, and hence the breaking index,
as is shown by the last equation, becomes greater or smaller than $3$
respectively. For example, if we consider the Crab pulsar, for which $\Omega
=190\;s^{-1}$, $\dot{\Omega}=-2.4\;10^{-9}s^{-2}$and $n=2.5$, they lead to:
\begin{eqnarray}
&&\frac{I_tc^3}{M^2\Omega ^2}\simeq 2\;10^3\,\sin ^2\theta \;yr\;, \\
&&\omega _f=1.2\left( 1+1.7\;10^{-3}\tan ^2\theta \right) s^{-1}\;, \\
&&\dot{\theta}\simeq 10^{-4}\;\tan \theta \;yr^{-1}\;.
\end{eqnarray}

The resulting value for $\omega_f$, of the order of seconds, is in good
agreement with the theoretical expectations and the bounds derived from the
Vela and the Crab pulsars. But this value for $\omega_f$ in fact corresponds
to a fine tuning to have $\frac{6\omega_fc}{5\Omega ^2R}-1\lesssim 10^{-3}$,
which is rapidly spoiled as $\Omega$ decreaces.
The effects of the
electromagnetic aligning torques are much higher than the corresponding ones
for a rigid star. In particular for a young star they give a characteristic
time of the order of 10 yr. For this reason the magnetic moment of a neutron
star
in the vacuum will reach the rotation axis or the equator, and stabilize in a
rather short time compared with the age of the star. If it tends to the
rotation axis the breaking index grows much bigger than 3, and if it falls in
the equatorial plane the breaking index becomes exactly 3. Clearly, this is
not the situation observed in the known pulsars.

The effects of the magnetosphere introduce a qualitative change in the
dipole alignment behavior. The lower magnetosphere can be modeled by the
Goldreich-Julian currents, in which case we can take $\nu _x=-1/4$. At the
moment there is no suitable calculation for the non anomalous torques,
generated by the upper magnetosphere, and thus we will maintain the
corresponding coefficients as phenomenological parameters. Then we have:
\begin{eqnarray}
\dot{\theta} &=&-\frac 23\frac{M^2\Omega^2}{I_c\,c^3}\left( \frac{3\omega _fc}
{10\Omega ^2R}+\nu _{yz}\right) \sin \theta \,\cos \theta \;, \\
\dot{\Omega} &=&-\frac 23\frac{M^2}{I_tc^3}\left( \nu _{yz}\sin
^2\theta +\tilde{\nu}_{yz}\right) \Omega ^3\;, \label{33}\\
n &=&3\left[ 1-\frac 13\frac{I_t}{I_c}
\left( \frac{3\omega _fc}{10\Omega ^2R}+\nu _{yz}\right) \sin \theta \,\cos
\theta \;\frac d{d\theta }\left( \nu _{yz}\sin ^2\theta +\tilde{\nu}
_{yz}\right) ^{-1}\right] \;.
\end{eqnarray}

There is a very important remark to be made about the $\theta $ behavior.
Taking into account that we have $-\frac{M^2\Omega^2}{I_tc^3}\simeq
\frac{\dot{\Omega}}\Omega $, the evolution of $\theta $ when $\frac{3
\omega _fc}{10\Omega ^2R}+\nu _{yz}\simeq 1$ is dominated by $\tan \theta
\propto e^{-\frac{\dot{\Omega}I_t}{\Omega I_c}t}$. For a typical neutron
star $\frac{{\Omega}I_c}{\dot\Omega I_t}\simeq 10^{-2}\tau_\Omega$, and
thus in a very short time the magnetic dipole lines up with the rotation
axis, and from there on $\dot{\theta}=0$. This behavior is a consequence of
the relative freedom of the crust respect to the core, which increases the
velocity of alignement by a factor $\frac{I_t}{I_c}$. But this is not
the only possible behavior. Another one can be realized if there is an
equilibrium point for the dynamics of $\theta $ at
\begin{equation}
\nu _{yz}(\theta )= -\frac{3\omega _f c}{10\Omega^2 R}\;,  \label{35}
\end{equation}
satifying $\nu _{yz}^{\prime }(\theta )\cos \theta >0$ to be stable, where
the prime indicates a derivative with respect to $\theta $. In this case $
\theta $ will rapidly adjust to the equilibrium value and its dynamics will
be tied to the angular velocity dynamics
\begin{equation}
\dot{\theta}=-\frac {3c}{10\nu_{yz}^{\prime }(\theta )R}\left(
\frac{\omega _f}{\Omega ^2}\right) ^{\prime }\dot{\Omega}\;,
\end{equation}
which implies that $\theta$ slides towards the rotation axis. Besides this,
from eq. (35) we have
\begin{equation}
\frac{\dot{\Omega}}\Omega \simeq \frac 23\frac{M^2}{I_tc^3}\left( \frac{
3\omega _fc}{10R}\sin ^2\theta -\tilde{\nu}_{yz}\Omega ^2\right) \;.
\end{equation}

If $\tilde{\nu}_{yz}\gtrsim \frac{3\omega _fc}{10\Omega ^2R}\sin ^2\theta
\simeq -\nu _{yz}\sin ^2\theta $, the angular velocity $\Omega $ decreases
throughout the life of the star; otherwise, if $\tilde{\nu}_{yz}<\frac{
3\omega _fc}{10\Omega ^2R}\sin ^2\theta $, the surroundings accelerates the
star. The first situation seems to apply to the known pulsars.

There are theoretical arguments suggesting that $\omega _f$ depends on $
\Omega ^k$, with $k<2$(\cite{ALS}; \cite{Mb}). The angular function
$\nu _{yz}(\theta )$ is of order one, but during the life of the star the
angular velocity constantly decreases. Thus, at a given moment the
equilibrium point condition (\ref{35}) cannot be maintained any more. When
$\frac{\omega _f}\Omega \frac c{\Omega R}$ becomes significantly greater than
$\nu _{yz}(\theta )$ there is a change of regime and the dipolar moment
rapidly aligns with the rotation axis. For a young star evolving at the
equilibrium point, where $\nu _{yz}$ is of order one, we can estimate a
lower bound for $\tau _f$ because it must be close to or greater than
$\frac {3c}{10\Omega ^2R}$. For example, for the Crab pulsar it is $\tau _f
\gtrsim \,0.25 s$ and for the Vela pulsar $\tau _f\gtrsim \,1.8 s$, consistent
with the known upper bound. In this regime the breaking index is given by
\begin{equation}
n-3=(k-2)\frac{\nu_{yz}}{\nu^\prime_{yz}}
\frac{\left(\nu_{yz}\sin^2\theta+\tilde\nu_{yz}\right)^\prime}
{\left(\nu_{yz}\sin^2\theta+\tilde\nu_{yz}\right)} \;,
\end{equation}
which has the correct order of magnitude. To have $n<3$ it must be
$\left(\nu_{yz}\sin^2\theta+\tilde\nu_{yz}\right)^\prime \cos\theta <0$
at the equilibrium point.

Although the aim of this model is to describe the overall evolution of the
magnetic dipole, a question that naturally arises is what it can say about
the glitches, which up to this point have not been considered. If we suppose
that when a glitch happens there is a change in $\theta$, from eq.
(\ref{33}) we have
\begin{equation}
\frac{\Delta \dot{\Omega}}{\dot{\Omega}}\simeq \frac{\left(\nu_{yz} \sin^2
\theta + \tilde\nu_{yz}\right) ^{\prime }}{\left(\nu_{yz} \sin^2
\theta + \tilde\nu_{yz}\right) }\Delta
\theta \;.
\end{equation}
The magnetosphere-dependent factor can be assumed
of the order of the unity, whereas the change in the angle $\theta$ is at
most of the same order as $\frac{\Delta
\Omega_c}{\Omega}\simeq \frac{I_t}{I_c}\frac{\Delta
\Omega}{\Omega}$, where $\Delta\Omega_c$ is the crust angular velocity
jump just after the glitch. Thus this relation implies that the relative
change of $\dot{\Omega}$ is at most $\frac{\Delta \dot{\Omega}}{\dot{\Omega}}
\simeq \frac{I_t}{I_c}\frac{\Delta \Omega }\Omega $. For example, in the
case of the Crab pulsar this
relation tells us that $\frac{\Delta \dot{\Omega}}{\dot{\Omega}}\simeq
10^2\frac{\Delta \Omega }\Omega \simeq 10^{-6}$, whereas for the observed
glitches it is $10^{-6}\lesssim \frac{\Delta \dot{\Omega}}{\dot{\Omega}}
\lesssim 10^{-3}$. This shows that these phenomena involve some non
systematic and probably very complicated factors, such as deformations,
earthquakes and breakings of the crust, which are not considered in the
present model. For this reason it gives only a lower boundary for the ratio
between the increment of the angular acceleration and the acceleration
itself. The change of the $\theta$ and $\omega_f$ parameters during the
glitches could produce a departure from the equilibrium point that would be
reached again after a transient time of the order of ten years.

\section{Concluding remarks}

In this paper we have developed a two-component model for the evolution of a
neutron star in presence of a magnetosphere. It considers the core and the
crust with a fixed magnetic dipole moment, taking into account the normal
and anomalous torques that act on the dipole and the friction between the
core and the crust. We also analize the effects of gravitational corrections
and show that they only introduce a renormalization in some parameters, but
do not affect the qualitative behavior of the system. We have solved the
equations of motion taken advantage of the very different characteristic
time scales that emerge from the complete dynamics.

The anomalous torques are usually supposed to be irrelevant for the
alignment of the magnetic dipole, but here we show that in fact they are
very relevant. This is because the crust-core interaction allows the angular
momentum of the crust to evolve independently of the magnetic moment by
interchanging angular momentum with the core. Despite this qualitative
change in the alignment, the energy loss due to dissipation is very small
and the equation governing the angular velocity is the same as for a rigid
star.

Another effect of the crust-core interaction is an amplification of the
classical alignment velocities by a factor $\frac{I_o}{I_c}$ that greatly
reduces the alignment times. This result and the observations lead us to the
conclusion that these torques are not directly governing the dynamics, but
there is an equilibrium point which effectively drives the angle
$\theta $. This equilibrium is reached by the interplay of the aligning
effect of the $\hat x$ axis torque, which depends on the friction between
core and crust, and the $\hat y-\hat z$ plane aligning torques, which
depends in a crucial way on the magnetosphere. The equilibrium point
naturally moves with the caracteristic times of the dynamics of the angular
velocity and leads to a breaking index near $3,$ as is actually observed.
This equilibrium point regime works for rapidly rotating young stars. When
the angular velocity becomes small it cannot be established and the magnetic
moment will rapidly reach its final state. The equilibrium point regime
requires a crust-core friction in reasonable agreement with the theoretical
expectations and the observational boundaries.

One can be tempted to extend this model to the study of the glitches, but
the results do not agree well with the observational data from the Crab
pulsar. We obtain only a
lower boundary for $\frac{\Delta \dot{\Omega}}{\dot{\Omega}}$. This is to be
expected, because in these phenomena it is very likely that more complicated
processes, such as crust deformation, fractures and earthquakes, not taken
into account here, have a very significant role. This suggests that other
effects could be included in this framework. For the electromagnetic
interaction, the actual effect of anomalous torques due to quadrupolar
magnetic moments should be investigated, although their average effect is
expected to be strongly suppressed in a rigid star model (\cite{GN}). The
effects due to crust elasticity and oblateness could also be important as
analized in references (\cite{G}; \cite{Ma}), and they might be relevant to
extend this model for describing the glitches dynamics.

In summary, this model gives a consistent description of the overall
evolution of the main parameters of a neutron star. As long as the magnetic
moment can be considered constant it is a suitable aproach for understanding
the magnetic field and angular velocity beheavior and, as has already been
said, it can be considered a first step for constructing a reliable
description of the global dynamics of a neutron star. In particular it
reconciles the fact that the angular velocity can be permanently decreasing
by radiation, and that simultaneously for a young and rapidly rotating star
we have a breaking index $n\lesssim 3$. For an older one the magnetic dipole
aligns with the rotation axis and stabilize.

\section{Acknowledgments}

This work has been realized with a partial support from the Consejo
Nacional de Investigaciones Cient\'{\i}ficas y T\'ecnicas (CONICET),
Argentina.

\figcaption{The crust angular velocity ${\bf\Omega_c}$ defines the z-axis.
The magnetic moment ${\bf M}$ is in y-z plane, with an angle $\theta$ with
respect to de z-direction. The core angular velocity ${\bf\Omega_o}$ forms
an angle $\alpha$ with ${\bf\Omega_c}$ and $\beta$ with ${\bf M}$.}


\begin{thebibliography}{}

\bibitem[Abney, Epstein \& Olinto 1996]{AEO} Abney, M., Epstein, R.I., \&
Olinto, A.V. 1996, ApJ, 466, L91
\bibitem[Alpar, Cheng \& Pines 1989]{ACP} Alpar, M.A., Cheng, K.S., \&
Pines, P. 1989, ApJ, 346, 823
\bibitem[Alpar, Langer \& Sauls 1984]{ALS} Alpar, M.A., Langer, S.A., \&
Sauls, J.A. 1984, ApJ, 282, 533
\bibitem[Alpar \& \"Ogelman 1990]{AO} Alpar, M.A., \& \"Ogelman, H. 1990,
ApJ, 349, L55
\bibitem[Alpar \& Sauls 1988]{AS} Alpar, M.A., \& Sauls, J.A. 1988, ApJ,
327, 725
\bibitem[Anderson \& Itoh 1975]{AI} Anderson, P.W., \& Itoh, N. 1975, Nature,
256, 25
\bibitem[Blandford, Applegate \& Hernquist 1983]{BAH} Blandford, R.D.,
Applegate, J.H., \& Hernquist, L. 1983, MNRAS, 204, 1025
\bibitem[Casini \& Montemayor 1997]{CM} Casini, H., \& Montemayor, R. 1997,
hep-th/9704053
\bibitem[Chanmugan 1992]{Ch} Chanmugan, G. 1992, ARA\&A, 30, 143
\bibitem[Chanmugam \& Sang 1989]{ChS} Chanmugam, G., \& Sang, Y. 1989, MNRAS,
241, 295
\bibitem[Davis \& Goldstein 1970]{DG} Davis, L., \& Goldstein, M. 1970, ApJ,
159, L81
\bibitem[Goldreich 1970]{G} Goldreich, P. 1970, ApJ, 160, L11
\bibitem[Goldreich \& Julian 1961]{GJ} Goldreich, P., \& Julian, W.H. 1961,
ApJ, 157, 869
\bibitem[Goldreich \& Reisenegger 1992]{GR} Goldreich, P., \& Reisenegger,
A. 1992, ApJ, 395, 250
\bibitem[Good \& Ng 1985]{GN} Good, M.L., \& Ng, K.K. 1985, ApJ, 299, 706
\bibitem[Gunn \& Ostriker 1969]{GO}  Gunn, J.E., \& Ostriker, J.P. 1969,
Nature, 221, 454
\bibitem[Lamb 1991]{L} Lamb, F. 1991, in ASP Conf. Proc. 20, {\sl Frontiers
of Stellar Evolution}, ed. D. Lambert (San Francisco: ASP), 299
\bibitem[Link \& Epstein 1996]{LEb} Link, B., \& Epstein, R.I. 1996, ApJ,
457, 844
\bibitem[Link \& Epstein 1997]{LEa} Link, B., \& Epstein, R.I. 1997, ApJ,
478, L91
\bibitem[Link, Epstein \& Baym 1992]{LEB} Link, B., Epstein, R.I., \& Baym,
G. 1992, ApJ, 390, L21
\bibitem[Lyne \& Manchester 1988]{LM} Lyne, A.G., \& Manchester, R.N.
1988, MNRAS, 234, 477
\bibitem[Manchester \& Taylor 1977]{MT} Manchester, R.N., \& Taylor,
J.H., 1977 {\sl Pulsars }(San Francisco: Freeman)
\bibitem[Macy 1973]{Ma} Macy, W.W.Jr. 1973, ApJ, 190, 153
\bibitem[Mc Cullock et al. 1990]{MHMK} Mc Cullock, P.M., Hamilton, P.A.,
Mc Connell, D., \& King, E.A. 1990, Nature, 346, 822
\bibitem[1991a, 1991b]{M1} Mendell, G. 1991, ApJ, 380, 515 ; ApJ,
380, 530
\bibitem[Mendell 1997]{Mb} Mendell, G. 1997, astro-ph/9702032
\bibitem[Michel 1983, 1985]{M} Michel, F.C. 1983, ApJ, 266, 288; 1985, ApJ,
290, 721
\bibitem[Michel \& Goldwire 1970]{MG} Michel, F.C., \& Goldwire, H.C. 1970,
Ap Letters, 5, 21
\bibitem[1995]{MP} Muslinov, A., \& Page, D. 1995, ApJ, 440,
L77
\bibitem[Pacini 1967, 1968]{P} Pacini, F. 1967, Nature, 216, 567 ;1968,
Nature, 219, 145
\bibitem[Pandey \& Prasad 1996]{PP} Pandey, U.S., \& Prasad, S.S. 1996, A\&A,
308, 507
\bibitem[Phynney \& Kulkarni 1994]{PK} Phynney, S., \& Kulkarni, S. 1994,
ARA\&A, 32, 591
\bibitem[1991a, 1991b, 1991c]{R} Ruderman, M. 1991, ApJ, 366, 261;
ApJ, 382, 576; ApJ, 382, 587
\bibitem[Ruderman, Zhu \& Chen 1997]{RZC} Ruderman, M., Zhu, T., \& Chen,
K. 1997
\bibitem[Srinivasan et al. 1990]{SBMT} Srinivasan, G., Bhattacharya, D.,
Muslinov, A., \& Tsygan, A. 1990, Current Science, 59, 31
\end{thebibliography}
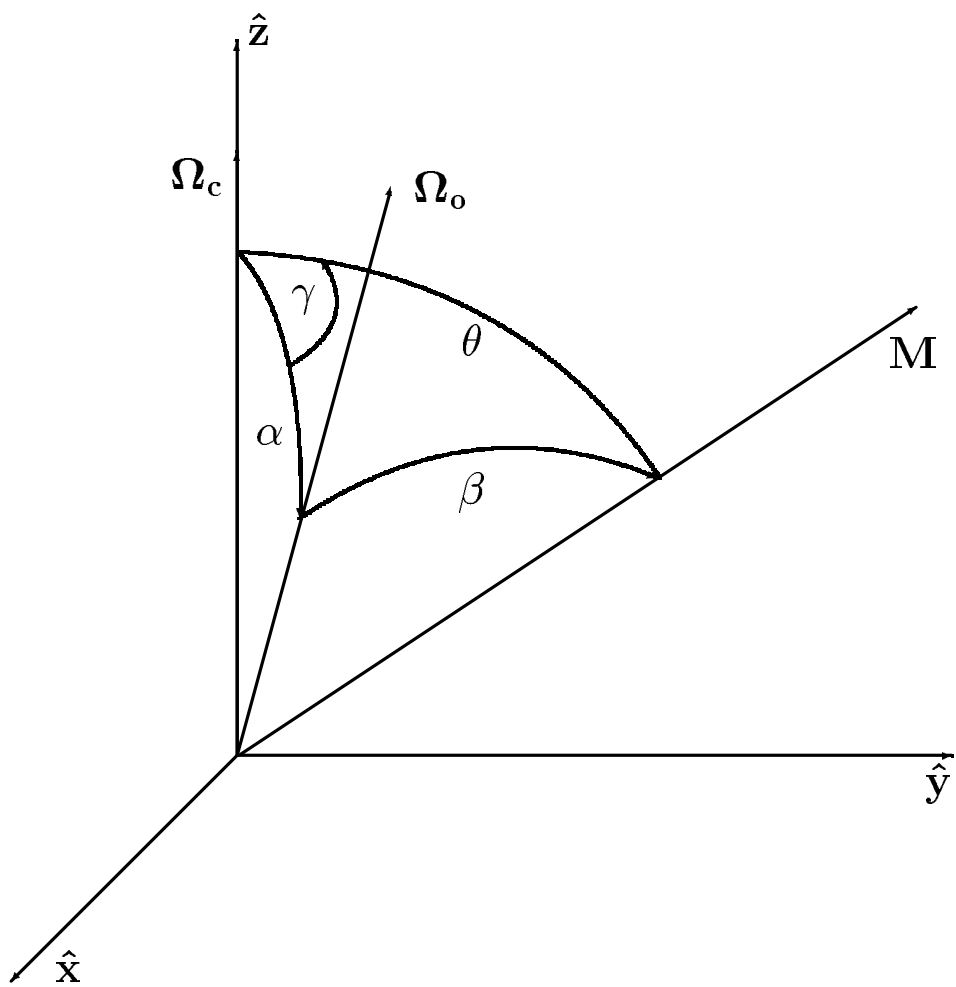
\end{document}